\documentclass[12pt]{article}

\usepackage{amsfonts,amssymb,amsmath,amsthm,graphicx, slashed, epsfig}

\newcommand{\be}{\begin{equation}}
\newcommand{\ee}{\end{equation}}
\newcommand{\bea}{\begin{eqnarray}}
\newcommand{\eea}{\end{eqnarray}}
\newcommand{\beq}{\begin{eqnarray}}
\newcommand{\eeq}{\end{eqnarray}}

\def\({\left(}
\def\){\right)}
\def\[{\left[}
\def\]{\right]}
\def\a{\alpha}

\begin{document}

\title{The Casimir effect: medium and geometry}

\author{Valery N. Marachevsky \thanks{email: maraval@mail.ru} \\
{\it Department of Theoretical Physics} \\ {\it Saint-Petersburg
State University}\\
{\it 198504 St. Petersburg, Russia.} }


\maketitle

\begin{abstract}
Theory of the Casimir effect is presented in several examples.
Casimir - Polder type formulas, Lifshitz theory and theory of the
Casimir effect for two gratings separated by a vacuum slit are
derived. Equations for the electromagnetic field in the presence of
a medium and dispersion are discussed. Casimir effect for systems
with a layer of $2+1$ fermions is studied.


\end{abstract}

\maketitle


\section{Introduction}

In the Casimir effect \cite{Casimir} one typically solves
electromagnetic or scalar classical boundary problems and finds
normal modes of the system.  The summation over eigenfrequencies of
normal modes of the system determines its ground state energy via a
relation $E=\sum_i \hbar \omega_i/2$. In the present paper this
relation for the ground state energy is applied to dielectrics or
metals separated by a vacuum slit, which is the typical system in
the Casimir effect. Interaction part of the Casimir energy of two or
more separated bodies in a vacuum is always finite, this part
determines Casimir forces measured in experiments.

Different geometries were studied in the theory of the Casimir
effect. In most geometries eigenfrequencies $\omega_i$ are not known
explicitly, in this case two main approaches exist to derive the
Casimir energy. One commonly used approach in the Casimir effect is
zeta function technique \cite{rev10, rev11}. An alternative
technique which proved to be efficient in the Casimir effect is the
scattering approach, it is discussed in Sec.$3$ with emphasis on
flat and periodic geometries.

Typically Casimir interaction is considered local in space with a
given frequency dispersion thus allowing nonlocality of photon
interaction in time direction. Spatial nonlocality of photon
interaction arising due to fermions and existence of polarization
diagrams in quantum electrodynamics make the issue technically more
complicated. Spatial dispersion is particularly important in metals,
it yields e.g. a Debye screening. Graphene is another example of
$2+1$ system where spatial dispersion can not be neglected.  A
complete theory which would give a description of $3+1$ systems with
spatial dispersion in the Casimir effect is still absent.

A problem of spatial dispersion is closely related to the high
temperature behavior of the Casimir force. The problem of the high
temperature asymptotics of the free energy between two metals
attracts attention of theoreticians and experimentalists.
Asymptotics of permittivity at small frequencies is important at
large separations between two metals or high temperatures. Two
models of metal permittivity, the plasma model and the Drude one,
yield high temperature asymptotics of free energy differing by a
factor of $2$, see Sec.$4$.

Instead of using a selected model of permittivity one can
alternatively evaluate components of the polarization operator in a
medium and write equations of motions. Components of the
polarization operator can be evaluated from the first principles
once one knows properties of quasiparticles in the medium. This
approach allows one to obtain high temperature behavior of the
system from the first principles of quantum electrodynamics and
quantum field theory (see Secs.$4$ and $5$). Spatial dispersion of
the polarization operator in graphene is important for correct
determination of high temperature behavior, see Sec.5 for details.

A layout of the
paper is the following. Sec.$2$ is devoted to derivation of the
Casimir-Polder interaction \cite{CasPol} of an anisotropic atom with
a flat surface (perfectly conducting and dielectric surfaces are
considered) on the basis of an approach developed in
Ref.\cite{MarPis}.

In Sec.$3$ the scattering theory approach to periodic systems is
developed after discussion of the Lifshitz formula for two parallel
flat surfaces separated by a vacuum slit $a$ \cite{Lifshitz}. Theory
of the Casimir effect for  periodic systems with frequency dependent
permittivity was developed in \cite{Mar2}. Comparison of theory and
experiments beyond applicability of the proximity force
approximation (PFA) was performed for grating geometries in
\cite{Mar2, Mar3}.

 In Sec.$4$  Maxwell equations
in a medium with a polarization operator of fermions taken into
account explicitly are discussed.  Free energy between two
superconductors is analyzed at large separations between
superconductors. Importance of nonlocality and spatial dispersion in
the Casimir effect is emphasized.

In Sec.$5$ fermions in a flat $2+1$ layer are studied following
ideas developed in Sec.$4$.  In particular, derivation of reflection
coefficients from a layer with $2+1$ fermions is given in Sec.$5$ in
a novel way, diagrams corresponding to the Casimir-Polder energy and
the Lifshitz free energy for systems with a flat $2+1$ fermion layer
are shown. Graphene is a typical example of such a system
\cite{Geim}, its properties in the Casimir effect have been studied
by different authors \cite{Mar1}-\cite{Gr7}. Finally the high
temperature asymptotics of a graphene - parallel ideal metal system
is derived.

Additional references for further reading are given after
Conclusions section.

We use coordinates $x^1, x^2, x^3$  and $x, y, z$ interchangeably.
The units $\hbar=c=k_B=1$ are used throughout the paper.

\section{Casimir-Polder energy}

We review general formalism developed in \cite{MarPis} and then
derive the Casimir-Polder energy of an anisotropic atom above a
perfectly conducting plane \cite{CasPol} and energy of an
anisotropic atom above a flat dielectric.

 The
atom is modeled as a localized electric dipole at the point
$(x^1,x^2,x^3)=(0,0,a)$, which is described by the current
$J_\mu(x)$:
\begin{align}
J_0 (x)&=
\sum_{i=1}^{3}d_{i}(t)\partial^{i}\delta(x^1)\delta(x^2)\delta(x^3-a),
\label{J0} \\ J_i(x)&=-
\dot{d}_{i}(t)\delta(x^1)\delta(x^2)\delta(x^3-a), \ i=1,2,3.
\label{Ji}
\end{align}
The condition of current conservation holds:
$$
\partial_{\mu}J^{\mu}=0,
$$
the expectation value of dipole moments is given by the formula
\cite{Berest}
 \begin{equation}
 \langle T (d_j(t_1)    d_k(t_2))  \rangle = - i
\int_{-\infty}^{+\infty} \frac{ e^{-i\omega(t_1-t_2)}}{2\pi} \,
\alpha_{jk} (\omega) \label{pol}d\omega ,
\end{equation}
where
 $\alpha_{jk} (\omega)$
for $\omega > 0$ coincides with atomic polarizability, a ground
state of the atom is represented by a bracket $\rangle$. We will use
the following expression for interaction energy $E$ of the atom with
a system $C$:
\begin{equation}
E = \lim_{T_1 \to +\infty} \left\langle\ln\left[\frac{\int
\exp\left\{i S(A)+i \int JA \, d^4 x\right\}DA}{\int \exp\left\{i
S(A)\right\}DA}\right] \right\rangle= \lim_{T_1 \to +\infty} - i
\frac{\langle J  D_C J \rangle}{2T_1}.
 \label{energia}
\end{equation}
$T_1$ is a time interval, $D_C$ is a propagator of photons in the
presence of the system $C$.

Consider Feynman gauge of vector potentials. A free photon
propagator in this gauge has the form
\begin{equation}
D^{(0)}_{\mu\nu} (t, r) = - i \int_{-\infty}^{+\infty}
\frac{d\omega}{2\pi} D^{(0)}_{\mu\nu} (\omega, r) e^{-i\omega t} ,
\end{equation}
where
\begin{equation}
D^{(0)}_{\mu\nu} (\omega, r) = - g_{\mu\nu} e^{i|\omega| r}/(4\pi r)
\label{vacprop}
\end{equation}
and $g_{\mu\nu}= (1, -1, -1, -1)$ (Heaviside - Lorentz units are
used). The propagator satisfying perfectly conducting boundary
conditions $A_0|_{z=0}=A_x|_{z=0}=A_y|_{z=0}=0$ and $\frac{\partial
A_z}{\partial z}|_{z=0}=0$  on a plane located at $z=0$ has the
form:
\begin{align}
D_{00} (\omega, r) &= - e^{i|\omega| r}/(4\pi r) + e^{i|\omega|
r_1}/(4\pi r_1) \nonumber
\\ D_{xx} (\omega, r) &= e^{i|\omega| r}/(4\pi r)
- e^{i|\omega| r_1}/(4\pi r_1) \label{Feynm}
\\ D_{yy} (\omega, r) &= e^{i|\omega| r}/(4\pi r) -
 e^{i|\omega| r_1}/(4\pi r_1)
\nonumber
\\ D_{zz} (\omega, r) &= e^{i|\omega| r}/(4\pi r) +
 e^{i|\omega| r_1}/(4\pi r_1)
, \nonumber
\end{align}
where $r=\sqrt{(x^\prime -x)^2 + (y^\prime - y)^2 + (z^\prime - z)^2
} $ and \newline $r_1=\sqrt{(x^\prime -x)^2 + (y^\prime - y)^2 +
(z^\prime + z)^2 } $. An image method yields opposite signs for
Dirichlet boundary conditions and equal signs for Neumann boundary
conditions in (\ref{Feynm}).

Performing integration over time, using (\ref{pol}),
(\ref{energia}), (\ref{Feynm}) and subtracting the contribution of
the free photon propagator (\ref{vacprop}) (to derive interaction
energy of the atom with a plane at $z=0$) one obtains after Wick
rotation the Casimir-Polder result \cite{CasPol} for energy of the
atom above a perfectly conducting plane:
\begin{align}
&E = - \int_{0}^{+\infty} \frac{d\omega}{2\pi} \sum_{i=1}^3
\alpha_{ii}(i\omega) \Bigl(-\omega^2 D_{ii} +
\frac{\partial^2}{\partial x_i \partial x_i^\prime}
D_{00}\Bigr)\Bigl|_{x^\prime=x, y^\prime=y, z^{\prime}=z=a} = \nonumber \\
&-\frac{1}{64 \pi^2 a^3} \int_{0}^{+\infty}d\omega \Bigl(
(\alpha_{11}(i\omega)+ \alpha_{22}(i\omega))(4\omega^2 a^2 +
2\omega a + 1)e^{-2\omega a}  + \\
&\quad\quad\quad\quad\quad\quad\quad\quad +\alpha_{33}(i\omega)
2(2\omega a + 1)e^{-2\omega a}\Bigr) \nonumber
\end{align}

To derive the result for the atom above a flat dielectric surface it
is convenient to apply the Weyl formula \cite{Weyl}. It is
instructive to solve the problem in Feynman gauge as well.
Components of the vector potential reflect from the surface in the
following way :
\begin{align}
A_p &\to  r_{TE} A_p \\
A_0 &\to -r_{TM} A_0 \\
A_l &\to -r_{TM} A_l \\
A_z &\to r_{TM} A_z,
 \end{align}
where $p$ is a transverse (perpendicular) direction and $l$ is a
longitudinal direction to the vector $(k_x, k_y)$, $r_{TE}(\omega,
k_x, k_y)$ is a reflection coefficient of the transverse electric
(TE) wave and $r_{TM}(\omega, k_x, k_y)$ is a reflection coefficient
of the transverse magnetic (TM) wave.

The free propagator has the same form (\ref{vacprop}). To make
needed decomposition of waves it is convenient to use the Weyl
formula:
\begin{align}
\frac{e^{i\omega r}}{4\pi r} &= \frac{i}{(2\pi)^2} \int\int e^{i(k_x
(x-x^\prime)+ k_y (y-y^\prime) +
\sqrt{\omega^2-k_x^2-k_y^2}(z-z^\prime))} \frac{dk_x
dk_y}{2\sqrt{\omega^2-k_x^2-k_y^2}}
\end{align}
valid for $z-z^\prime>0$. The atom is located at the point
$(x,y,z=a)$, the surface of a dielectric is at $z=0$. According to
the Weyl formula the propagator is decomposed into infinite number
of plane waves having the wave vector $(k_x, k_y, \sqrt{\omega^2-
k_x^2- k_y^2})$. A projection of $A_x$ perpendicular to the
incidence plane is $A_x \cos(\theta) \exp(-ik_z z^\prime)$ ($\theta=
\arctan(k_y/k_x)$). After reflection from the surface it is
transformed into $A_x r_{TE} \cos(\theta) \exp(i k_z z^\prime) $,
and the projection of the reflected wave from the surface back to
the $x$-axis yields the factor $A_x r_{TE} \cos^2(\theta) \exp(i k_z
z^\prime) $. The projection of $A_x$ to the incidence plane and back
to the $x$-axis yields the factor $-A_x r_{TM} \sin^2(\theta) \exp(i
k_z z^\prime)$. Thus for the reflected from the surface vector
potential $A_x$ one gets:
\begin{align}
A_x=\frac{i}{(2\pi)^2} \iint_{-\infty}^{+\infty} dk_x dk_y
&\frac{e^{i(k_x (x-x^\prime)+ k_y (y-y^\prime) +
\sqrt{\omega^2-k_x^2-k_y^2}(z+z^\prime))}}{2\sqrt{\omega^2-k_x^2-k_y^2}
}\times \\ &\times \bigl(r_{TE}\cos^2(\theta)- r_{TM}\sin^2(\theta)
\bigr) \nonumber
\end{align}
Putting $x=x^\prime, y=y^\prime, z=z^\prime=a$ it is convenient to
perform integration over $\theta$ which yields $\pi$. Performing the
Wick rotation and assuming reflection coefficients depend on
$k=\sqrt{k_x^2 + k_y^2}$ one obtains the contribution of $D_{xx}$ to
the Casimir-Polder energy:
\begin{align}
&\frac{1}{(2\pi)^2} \int_0^{+\infty} d\omega \alpha_{xx}(i\omega)
\omega^2 \int_0^{+\infty} dk k \frac{ r_{TE}(i\omega,k) - r_{TM}
(i\omega, k)}{4\sqrt{\omega^2+k^2}} e^{-2a\sqrt{\omega^2+k^2}}
\end{align}
Similar steps allow one to obtain the following resulting expression
for the Casimir-Polder energy of an anisotropic atom above a
dielectric:
\begin{align}
E=& \frac{1}{(2\pi)^2} \int_0^{+\infty} d\omega
(\alpha_{xx}(i\omega) +\alpha_{yy}(i\omega)) \times \nonumber
\\ &\times \int_0^{+\infty} dk k \frac{ \omega^2
(r_{TE}(i\omega,k) - r_{TM} (i\omega,k)) - r_{TM}(i\omega, k) k^2
}{4\sqrt{\omega^2+k^2}} e^{-2a\sqrt{\omega^2+k^2}} -
\nonumber \\
&-\frac{1}{(2\pi)^2} \int_0^{+\infty} d\omega \alpha_{zz}(i\omega)
\int_0^{+\infty} dk k \frac{r_{TM}(i\omega, k)
k^2}{2\sqrt{\omega^2+k^2}} e^{-2a\sqrt{\omega^2+k^2}} \label{anisot}
\end{align}

Note that in the case of isotropic polarizability of the atom in the
$xy$ plane $\alpha_{I}(i\omega)$ the expression (\ref{anisot}) can
be rewritten in the following form:
\begin{multline}
E=  \int_0^{+\infty} d\omega \int_0^{+\infty} dk k
\Bigl(\Pi_{ll}^{at}(i\omega) D_{ll}(i\omega, k) +
\Pi_{pp}^{at}(i\omega) D_{pp}(i\omega, k) + \\
+\Pi_{zz}^{at}(i\omega) D_{zz}(i\omega,k)\Bigr), \label{an}
\end{multline}
where
\begin{align}
D_{ll}(i\omega, k) &= \frac{1}{(2\pi)^2} \frac{-r_{TM}(i\omega,
k)\sqrt{\omega^2
+k^2}e^{-2 a \sqrt{\omega^2+k^2}} } {2\omega^2}  , \\
D_{pp}(i\omega, k) &= \frac{1}{(2\pi)^2}\frac{r_{TE}(i\omega,
k)e^{-2 a
\sqrt{\omega^2+k^2}} }{2\sqrt{\omega^2+k^2}} ,\\
D_{zz}(i\omega, k) &= \frac{1}{(2\pi)^2}\frac{- r_{TM}(i\omega, k)
k^2 e^{-2 a \sqrt{\omega^2+k^2}} }{ 2 \omega^2 \sqrt{\omega^2+k^2} }
,
\end{align}
and $\Pi_{ll}^{at}(i\omega) = \Pi_{pp}^{at}(i\omega) =
\alpha_{I}(i\omega) \omega^2$, $\Pi_{zz}^{at}(i\omega)=
\alpha_{zz}(i\omega) \omega^2$.


 The formula (\ref{an}) can also be derived by
changing the basis locally from the beginning using in every point
of the momentum space $(k_x, k_y)$ in the Weyl formula the following
identity for a local basis change in two dimensions:
$\delta_{\mu\nu} = e_{x\mu} e_{x\nu} + e_{y\mu} e_{y\nu} = e_{l\mu}
e_{l\nu} + e_{p\mu} e_{p\nu} $ ($e_x, e_y$ and $e_l, e_p$ are two
sets of orthonormal vectors), so that the new orthonormal frame is
written in coordinates $l, p$, and respective components of the
photon propagator $D_{ll}, D_{pp}$ and the polarization operator
$\Pi_{ll}^{at}, \Pi_{pp}^{at}$ are written in this local frame in
momentum space.

\section{Free energy}
Consider two bodies separated by a vacuum slit.
To obtain the Casimir energy and free energy one has to determine
eigenfrequencies of normal modes of the electromagnetic field.
Eigenfrequencies of normal modes can be summed up by making use of
the argument principle \cite{Schram, Barash}, which states:
\begin{equation}
\frac{1}{2\pi i} \oint \phi(\omega) \frac{d}{d\omega} \ln f(\omega,
\beta) d\omega = \sum \phi (\omega_i) -\sum \phi(\omega_\infty) ,
\label{arg}
\end{equation}
where $\omega_i$ are zeroes and $\omega_\infty$ are poles of the
function $f(\omega, \beta)$ inside the contour of integration,
degenerate eigenvalues are summed according to their multiplicities,
also we assume that eigenfrequencies $\omega_i$ may depend on
continuous variables $\beta$. The equation for eigenfrequencies
$\omega_i$ of the corresponding problem of classical electrodynamics
is $f(\omega_i, \beta)=0$. For the Casimir energy  $\phi(\omega) =
\omega/2$.

In the absence of dissipation (when eigenfrequencies of Maxwell
equations are real) the contour of integration  in $\omega$ plane in
(\ref{arg}) first passes around eigenfrequencies $\omega_i$ and
branch cuts on the positive real frequency axis as it is explained
in detail e.g. in Ref.\cite{Nesterenko}. To obtain free energy one
has to substitute $\phi(\omega)= T\ln (2\sinh(\omega/2T))$ into
(\ref{arg}). Then free energy can be written as
\begin{align}
 {\mathcal F} =& -\frac{T}{\pi}\int d\beta \int_0^{+\infty} d\omega \ln (2\sinh(\omega/2T))
{\rm Im} \frac{\partial}{\partial \omega} \ln f(\omega, \beta) = \nonumber \\
=& \frac{1}{2\pi}\int d\beta \int_{0}^{+\infty} d\omega
\coth(\omega/(2T)) {\rm Im}\ln f(\omega, \beta) . \label{free3}
\end{align}
 Using the condition $ f(-\omega^*, \beta) = f(\omega, \beta)^*$
 one expands the contour of integration above
 the real frequency axis: $\omega \in (-\infty + i\varepsilon \ldots
 +\infty +i\varepsilon)$ and adds
 to this contour a semicircle integral
 around zero frequency in the opposite direction
(this semicircle integral is a contribution of the zero frequency
Matsubara term). The contour above the real frequency axis is moved
around the positive imaginary frequency axis, the poles of
$\coth(\omega/(2T))$ (with residues $2T$) yield Matsubara
frequencies $\omega_n = 2\pi n T$, and free energy is rewritten in
the form
\begin{equation}
{\mathcal F} = T \int d\beta \, {\sum_{n=0}^{\infty}}^\prime \ln f(i
\omega_n, \beta) , \label{free7}
\end{equation}
 prime means $n=0$
term is taken with the coefficient $1/2$. (Branch cuts on the real
frequency axis disappear once one puts perfectly conducting plates
at large separations. Scattering of electromagnetic waves between
these perfectly conducting plates yields eigenfrequencies $\omega_l$
located on positions of branch cuts. Eigenfrequencies $\omega_l$
effectively transform into eigenfrequencies of scattering states
which form branch cuts when perfectly conducting plates are moved to
spatial infinity.) In the presence of dissipation one can also use
formulas (\ref{free3}), (\ref{free7}), see a review \cite{Barash}.

Consider plane-plane geometry when two dielectric parallel slabs
(slab $1$: $z<0$, slab $2$: $z>a$) are separated by a vacuum slit
($0<z<a$) following Ref.\cite{Mar2}. In this case $TE$ and $TM$
modes are not coupled. The equation for $TE$ eigenfrequencies is
$f(\omega_i, k_x, k_y)=0$, where
\begin{equation}
f(\omega, \beta = (k_x, k_y))= 1-r^{(1)}_{TE down} (k_x, k_y,
\omega) r^{(2)}_{TE up} (k_x, k_y, \omega, a). \label{char}
\end{equation}
Here $r^{(1)}_{TE down}(k_x, k_y, \omega)$ is  the reflection
coefficient of a downward plane wave which reflects on a dielectric
surface of a slab $1$ at $z=0$, while $r^{(2)}_{TE up} (k_x, k_y,
\omega, L)$ is the reflection coefficient of an upward plane wave
which reflects on a dielectric surface of a slab $2$ at $z=a$. One
can deduce from Maxwell equations that $r^{(2)}_{TE up}(k_x, k_y,
\omega, L)= r^{(2)}_{TE down}(k_x, k_y, \omega) \exp(2ik_z a)$
($r^{(2)}_{TE down}(k_x, k_y, \omega)$ is a reflection coefficient
of a downward $TE$ plane wave which reflects from a dielectric slab
$2$ temporarily located at the position of the slab $1$, i.e. at
$z<0$).

After substitution of (\ref{char}) into the formula (\ref{free7})
one gets $TE$ part contribution to  free energy of two parallel
plates. Contribution of $TM$ part is obtained in full analogy. Free
energy has the form
\begin{equation}
    {\mathcal F}
    =T{\sum_{n=0}^{\infty}}^\prime\iint\frac{dk_x dk_y}{(2\pi)^2} \ln [(1-e^{-2a\sqrt{\omega_n^2+ k^2}}r_{TE}^{(1)}r_{TE}^{(2)})
        (1-e^{-2a\sqrt{\omega_n^2+ k^2}}r_{TM}^{(1)}r_{TM}^{(2)})] ,
        \label{EL}
\end{equation}
$\omega_n=2\pi n T$ are Matsubara frequencies, prime means $n=0$
term is taken with the coefficient $1/2$, $r^{(1)}_{TE} \equiv
r^{(1)}_{TEdown}, r^{(2)}_{TE} \equiv r^{(2)}_{TEdown}$, reflection
coefficients are evaluated at Matsubara frequencies.


Reflection coefficients (Fresnel coefficients) for  transverse
magnetic and electric flat waves approaching from vacuum the flat
surface of a medium described by dielectric permittivity
$\varepsilon(\omega)$ are well known:
\begin{align}
r_{TM}(\omega, k) &= \frac{\varepsilon(\omega) k_z^{(v1)} -
k_z^{(m1)} }
{\varepsilon(\omega) k_z^{(v1)} + k_z^{(m1)} },  \label{r1}\\
r_{TE}(\omega, k) &= \frac{k_z^{(v1)} - k_z^{(m1)}}{k_z^{(v1)} +
k_z^{(m1)} } , \label{r2}
\end{align}
where
\begin{align}
k_z^{(v1)} &= \sqrt{\omega^2 - k_x^2 -k_y^2} , \\
k_z^{(m1)} &= \sqrt{\varepsilon(\omega)\, \omega^2 - k_x^2 - k_y^2}
.
\end{align}
The Lifshitz result for two parallel semi-infinite dielectrics
separated by a vacuum slit is obtained from (\ref{EL}) when Fresnel
coefficients $r_{TM}(i\omega, k), r_{TE}(i\omega, k) $ are
substituted into (\ref{EL}) \cite{Lifshitz, Kats}.

The limit $\varepsilon(i\omega) \to + \infty$ corresponds to
reflection coefficients of a perfectly conducting metal plane:
$r_{TM}=+1, r_{TE}=-1$. In this case  one gets from (\ref{EL})
energy of two parallel perfectly conducting plates at zero
temperature, the result by Casimir \cite{Casimir}:
\begin{equation}
E =  \iiint \frac{d\omega d k_x d k_y}{(2\pi)^3} \ln(1-
e^{-2a\sqrt{\omega^2+k_x^2+k_y^2}}) = - \frac{\pi^2}{720 a^3}.
\end{equation}

Another limit is the high temperature asymptotics ($4\pi Ta\gg 1$)
of the free energy for two parallel perfectly conducting plates. It
can be obtained by evaluating contribution of the zero frequency
Matsubara term in (\ref{EL}):
\begin{equation}
{\mathcal F}|_{4\pi Ta\gg1} = T \iint \frac{d k_x d k_y}{(2\pi)^2}
\ln (1- e^{-2a\sqrt{k_x^2+k_y^2}}) = - \frac{T \zeta(3)}{8 \pi a^2}.
\label{ideal}
\end{equation}
We will continue discussion of high temperature behavior of free
energy for models of metals with frequency dispersion of
permittivity and for graphene systems in Sections $4$ and $5$.

\begin{figure}
\centering \includegraphics[width=6cm]{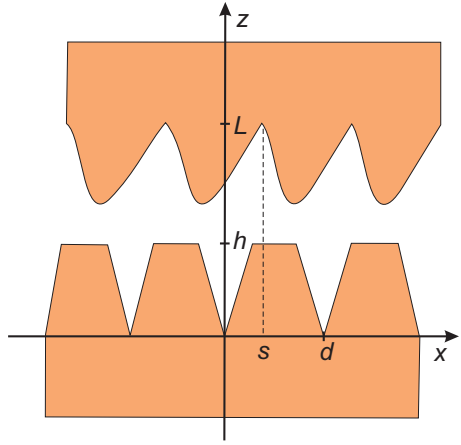} \caption{Two
gratings with a coinciding period $d$.} \label{grat14}
\end{figure}

Consider now the system of two periodic gratings with a coinciding
period $d$ separated by a vacuum slit (Fig.\ref{grat14})
\cite{Mar2}. For this system  one has to consider reflection of
downward and upward waves from a unit cell $-\pi/d<k_x<\pi/d$.
Imagine we remove the upper grating from the system. The reflection
matrix of the downward wave is defined as $R_{1 down}$ then.

The solution of Maxwell equations with a given $m$ ($m$ is an
integer number determining the Brillouin zone of the downward wave
with given $\omega, k_x, k_y$) for longitudinal components of the
electromagnetic field outside the corrugated region ($z \ge h$) may
be written by making use of the Rayleigh expansion \cite{Rayleigh}
for an incident monochromatic wave :
 \begin{align}
E_y(x,z, m) =& I_m^{(E)} \exp (i \alpha_m x - i \beta_m z)  +
\nonumber \\
 & \sum_{n=-\infty}^{+\infty}
R_{nm}^{(E)} \exp( i \alpha_n x  + i \beta_n z)  , \label{Ezp} \\
B_y(x,z, m) =& I_m^{(B)} \exp (i \alpha_m x - i \beta_m z)  +
\nonumber \\
 &\sum_{n=-\infty}^{+\infty} R_{nm}^{(B)} \exp( i
\alpha_n x  + i \beta_n z)  , \label{Hzp}\\
\alpha_n =& k_x + 2\pi n/ d  ,\quad \beta_n^2 = \omega^2 - k_y^2 -
\alpha_n^2     .
\end{align}
This solution is valid outside any periodic structure in $x$
direction, in our notations it is valid for $z \ge h$. All other
field components can be expressed in terms of longitudinal
components $E_y, B_y$ by standard formulas in waveguide theory. This
can be done since the factor $\exp(i k_y y)$ is conserved after
reflection of the electromagnetic wave from a grating.

To construct the reflection matrix one has to find Rayleigh
expansions with the condition $I_m^{(E)}=0, I_m^{(B)}=1$ and with
the condition $I_m^{(E)}=1, I_m^{(B)}=0$, $m=-\infty \ldots +
\infty$, $m$ is an integer number. For actual calculations one puts
$m=-J\ldots J$, where $J$ is an upper and $-J$ is the lower limit in
the sums (\ref{Ezp}), (\ref{Hzp}). In the case of reflection from a
grating the reflection matrix is given by
\begin{multline}
R_{1 down}(k_x, k_y, \omega) = \\ =\begin{pmatrix} R_{n_1
q_1}^{(E)}(I_m^{(E)}=\delta_{m q_1}, I_m^{(B)}=0  ) \qquad
&R_{n_2 q_2}^{(E)}(I_m^{(E)}=0, I_m^{(B)}=\delta_{m q_2} )   \\
R_{n_3 q_3}^{(B)}(I_m^{(E)}=\delta_{m q_3}, I_m^{(B)}=0  ) \qquad
&R_{n_4 q_4}^{(B)}(I_m^{(E)}=0, I_m^{(B)}=\delta_{m q_4} )
\end{pmatrix} . \label{R1}
\end{multline}

Imagine now that we remove the lower grating from the system (see
Fig.\ref{grat17}). We denote the reflection matrix of the upward
wave as $R_{2 up}$ then. Reflection matrices $R_{1 down}, R_{2up}$
depend on wave vectors of incident waves, parameters of gratings and
mutual location of the gratings. The equation for normal modes
states:
\begin{equation}
R_{1 down} (k_x, k_y, \omega_i) R_{2up}(k_x, k_y, \omega_i, L, s)
\psi_i = \psi_i, \label{bound}
\end{equation}
where $\psi_i$ is an eigenvector describing the normal mode with a
frequency $\omega_i$.
 Instead of
equation (\ref{char}) one obtains:
\begin{equation}
f_2(\omega, \beta=(k_x, k_y) ) = \det(I - R_{1 down} (k_x, k_y,
\omega) R_{2 up}(k_x, k_y, \omega, L, s)) .\label{EIG}
\end{equation}
For every $k_x, k_y$ the solution of $f_2(\omega_i)=0$ yields
possible eigenfrequencies $\omega_i$ of normal mode solutions of
Maxwell equations.

\begin{figure}
\centering \includegraphics[width=6cm]{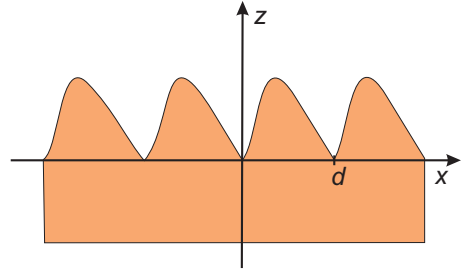} \caption{A
fictitious grating for which one evaluates $R_{2 down}$.}
\label{grat16}
\end{figure}
\begin{figure}
\centering \includegraphics[width=6cm]{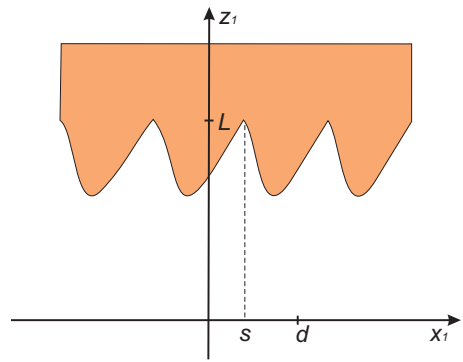} \caption{The
upper grating in Fig.\ref{grat14} for which one evaluates $R_{2up}$,
normal and lateral displacements from the fictitious grating shown
in Fig.\ref{grat16} are denoted by $L$ and $s$ respectively.}
\label{grat17}
\end{figure}
Suppose that the reflection matrix $R_{2 down}$ for a reflection
from the fictitious imaginary grating located as in Fig.\ref{grat16}
is known in coordinates $(x,z)$.
 Performing a
change of coordinates $z= - z_1 +L$, $x=x_1 - s \quad (s < d)$ in
$(\ref{Ezp}), (\ref{Hzp})$, it is possible to obtain a matrix $R_{2
up}$ for reflection of upward waves from a grating with the same
profile turned upside-down, displaced from the lower grating by
$\Delta x=s,\, \Delta z=L$ (see Fig.\ref{grat17}). It follows that
\begin{multline}
R_{2 up}(k_x, k_y, i\omega, L, s)= \\ = Q^*(s) K(k_x, k_y, i\omega,
L) R_{2 down}(k_x, k_y, i\omega) K(k_x, k_y, i\omega, L) Q(s) ,
\label{RR}
\end{multline}
where $R_{2 down}(k_x, k_y, i\omega)$ is a reflection matrix of
downward waves from the grating in the system of coordinates $(x,z)$
depicted on Fig.\ref{grat16}. Here $K(k_x, k_y, i\omega, L)$ is a
diagonal $2(2J+1)$ matrix of the form:
\begin{equation}
K(k_x, k_y, i\omega, L) = \begin{pmatrix} G_1 &  0  \\
0  &  G_1
\end{pmatrix},  \label{K}
\end{equation}
with matrix elements $e^{-L\sqrt{\omega^2+k_y^2+(k_x+\frac{2\pi
m}{d})^2}}$ on a main diagonal of a matrix $G_1$, $\quad m=-J\ldots
J$. Due to exponential factors in (\ref{K}) all resulting
expressions are finite. The lateral translation $2(2J+1)$ diagonal
matrix $Q(s)$ is defined as follows:
\begin{equation}
Q(s) = \begin{pmatrix}
G_2 &  0  \\
0  &  G_2
\end{pmatrix} ,
\end{equation}
with matrix elements $e^{2\pi i m s / d} \quad (m=-J\ldots J)$ on a
main diagonal of the matrix $G_2$.

 The summation over eigenfrequencies is
performed by making use of the formula (\ref{arg}), which yields the
Casimir energy of two parallel gratings on a unit surface:
\begin{multline}
E = \frac{1}{(2 \pi)^3} \int_0^{+\infty} d \omega
\int_{-\infty}^{+\infty} d k_y   \int_{-\frac{\pi}{d}}^{\frac{\pi}{d}} d k_x \\
\times \ln {\rm det} \Bigl(I - R_{1 down} (k_x, k_y, i\omega)
R_{2up}(k_x, k_y, i \omega, L, \varphi) \Bigr) , \label{EC}
\end{multline}
here $\varphi = 2\pi s/ d $, $s$ is a lateral displacement of two
gratings. This is an exact expression valid at zero temperature for
two arbitrary parallel gratings with coinciding periods $d$
separated by a vacuum slit.


Free energy on a unit surface $\mathcal{F}$ in the system of two
gratings can be written as follows:
\begin{multline}
\mathcal{F} (L, \varphi) = \frac{T}{\pi^2} \sum_{n=0}^{+\infty \:
\prime} \int_0^{+\infty} dk_y \int_0^{\pi/d} dk_x \\ \times \ln {\rm
det} \Bigl(I - R_{1 down} (k_x, k_y, i\omega_n) R_{2 up}(k_x, k_y,
i\omega_n, L, \varphi) \Bigr), \label{Free}
\end{multline}
here $\omega_n= 2\pi n T $ is a Matsubara frequency. The $n=0$ term
is multiplied by $1/2$. The formula (\ref{Free}) is valid for
arbitrary profile and arbitrary dielectric permittivity of each
grating.

\section{Polarization}

Interaction of the electromagnetic potential $A_{\mu}$ with a
current $J$ given by (\ref{J0}), (\ref{Ji}) yields formalism with
frequency dependent dielectric permittivity $\varepsilon(\omega)$.
This interaction is local and gauge invariant. Polarization bubbles
make interaction of photons in a medium essentially nonlocal.
Nonlocality in time is reflected in the frequency dependent
permittivity $\varepsilon(\omega)$.  More general nonlocality leads
to spatial dispersion, which by now led to intensive discussions
\cite{Dis1} - \cite{Pit3}.

 It
is instructive to write equations of motion in the presence of a
medium with a polarization operator taken into account:
\begin{equation}
\partial_{\nu} F^{\mu \nu} + \Pi^{\mu \nu} A_{\nu} = -j^{\mu},
\end{equation}
$j^{\mu}$ is an external current, $\Pi^{\mu\nu}$ is a polarization
operator.

In momentum space equations of motion can be written as follows:
\begin{align}
{\rm div} \vec{E} + \frac{\Pi^{0n} E^n}{i\omega} &= j^0(\omega,
\vec{k}) \label{e1}\\
 -\omega^2 E^m + ({\rm rot} {\rm rot} \vec{E})^m + \Pi^{m
n} E^n &= i\omega j^m(\omega, \vec{k}) \label{e2}
\end{align}
Here ${\rm div}\vec{E} \equiv i k^n E^n$, $({\rm rotrot} \vec{E})^m
= (\vec{k}^2 \delta^{mn} - k^m k^n) E^n$.

 The components of the $3+1$
polarization operator can be evaluated using techniques of quantum
electrodynamics in condensed matter. General structure of the
polarization operator  resulting from gauge invariance in the
presence of a medium is the following \cite{Rubakov} (there is no
Lorentz invariance in the presence of a medium, only rotation
symmetry is conserved):
\begin{align}
\Pi^{0n}(\omega, \vec{k}, T) &=  \frac{\omega k^n \Pi^{00}(\omega, \vec{k}, T)}{\vec{k}^2} \\
\Pi^{mn}(\omega, \vec{k}, T) &= \frac{\omega^2 k^m k^n
\Pi^{00}(\omega, \vec{k}, T)}{\vec{k}^4} + \biggl(\delta^{mn} -
\frac{k^m k^n}{\vec{k}^2}\biggr) \Pi^{tr}(\omega, \vec{k}, T)
\end{align}
Due to these properties the equation (\ref{e1}) can be rewritten as
\begin{equation}
({\rm div}\vec{E}) \Biggl(1- \frac{\Pi^{00}(\omega,
\vec{k})}{\vec{k}^2}\Biggr) =
 j^0(\omega, \vec{k}) .
\end{equation}
In this approach the longitudinal dielectric permittivity is
naturally defined as $\varepsilon_l(\omega, \vec{k}) = 1 -
\Pi^{00}(\omega, \vec{k})/\vec{k}^2$, which coincides with random
phase approximation (RPA) result for the longitudinal dielectric
permittivity \cite{Lindhard}.

In equation (\ref{e2}) the part $\Pi^{m n} E^n$ can be expanded in
powers of $\vec{k}^2$. The leading term of this expansion is
\begin{align}
 &\Pi^{mn} E^n|_{|\vec{k}|\to 0} = -\frac{k^m (\vec{k}
\vec{E})}{\vec{k}^2} (\varepsilon_l(\omega, \vec{k},
 T)_{|\vec{k}|\to 0} - 1) \omega^2 - \label{term1}
\\ &-\biggl(E^m - \frac{k^m (\vec{k}\vec{E})}{\vec{k}^2}\biggr)
(\varepsilon_{tr}(\omega, \vec{k}, T)|_{|\vec{k}|\to 0} -1) \omega^2
, \nonumber
\end{align}
where
\begin{align}
\varepsilon_l(\omega, \vec{k}, T) &\equiv 1 - \Biggl(
\frac{\Pi^{00}(\omega, \vec{k}, T)}{\vec{k}^2} \Biggr) \\
\varepsilon_{tr}(\omega, \vec{k}, T) &\equiv 1 -
\frac{\Pi^{tr}(\omega, \vec{k}, T)}{\omega^2}
\end{align}
are frequency, momentum and temperature dependent longitudinal and
transverse dielectric permittivities.

In the limit of zero temperature $T\to 0$ the longitudinal and the
transverse dielectric permittivities coincide for $|\vec{k}|\to 0$,
$\varepsilon_l(\omega, \vec{k}, T)|_{|\vec{k}|\to 0, T\to 0} =
\varepsilon_{tr}(\omega, \vec{k}, T)|_{|\vec{k}|\to 0, T\to 0}
\equiv \varepsilon(\omega) $, and the term (\ref{term1}) results in
\begin{equation}
\Pi^{mn} E^n|_{|\vec{k}|\to 0, T\to 0} = -(\varepsilon(\omega)-1)
E^m . \label{term2}
\end{equation}
The term (\ref{term2}) is a standard one in the theory of a medium
without spatial dispersion.

Neglecting  dependence on $\vec{k}$ in the polarization operator we
get the standard equation for propagation of waves in a dielectric
medium with frequency dependent dielectric permittivity
$\varepsilon(\omega)$:
\begin{equation}
- \omega^2 \varepsilon(\omega) E^m + ({\rm rotrot} \vec{E})^m =
i\omega j^m (\omega, \vec{k}) , \label{main}
\end{equation}
The equation in a medium  has the same form (\ref{main}) in the
coordinate space, where the operation ${\rm rot}$ in (\ref{main}) is
a standard one in the coordinate space.

At $T=0$ one gets $\lim_{|\vec{k}| \to 0} \Pi^{00}(\omega,
\vec{k})/|\vec{k}|^2 = \omega_{pl}^2/\omega^2$ already for
non-relativistic electron gas, $\omega_{pl}^2 = 4\pi n_0 e^2 /m$,
$\omega_{pl}$ is a plasma frequency, $n_0$ is a charge density, $m$
is an electron mass. So general scheme outlined in this part of the
paper naturally yields Lindhard RPA result for
$\varepsilon_l(\omega, \vec{k})$ \cite{Lindhard} and the plasma
model of dielectric permittivity in the limit $|\vec{k}|\to 0$ of
the Lindhard formula: $\varepsilon_l(\omega, \vec{k})|_{|\vec{k}|
\to 0} = \varepsilon(\omega) = 1 - \omega_{pl}^2/\omega^2$.

Several comments must be maid about the plasma - Drude high
temperature problem in the Casimir effect. Permittivity in metals at
small frequencies is a Drude one ($\varepsilon_{Drude}(\omega) = 1 -
\omega_{pl}^2/\omega(\omega + i\gamma)$) because the current is
proportional to the electric field $\vec{E}$, and as a result the
pole of the first order is required in the permittivity.  Optical
data for metal permittivities $\varepsilon(\omega)$ at small
frequencies are in a good agreement with a Drude model of the
permittivity \cite{Palik}. Due to this it is challenging to explain
results of measurements of the Casimir force between two metals at
$T=300$K which were found to be in agreement with a plasma model of
the permittivity \cite{Decca}.

The main difference of plasma - Drude results follows from the zero
frequency $TE$ Matsubara term in free energy. Zero frequency
Matsubara term determines the asymptotics of the free energy at
large separations $4\pi T a \gg 1$. The Drude model predicts
$r_{TE}(i\omega=0,k)=0$ ($\lim_{\omega \to 0} \varepsilon(i\omega)
\omega^2 = 0$) and $r_{TM}(i\omega=0, k)=1$, while the plasma model
predicts $r_{TE}(i\omega=0, k) \ne 0$ ($\lim_{\omega \to 0}
\varepsilon(i\omega) \omega^2 = \omega_{pl}^2$) and
$r_{TM}(i\omega=0, k)=1$. At large separations between two metals
the zero frequency Matsubara term yields leading contribution to
free energy, and the asymptotical result for a plasma model of
permittivity is two times larger than the result for a Drude model
of permittivity.

In superconductors the current is proportional to vector potential
$A_{\mu}$ (or, better to say, it is proportional to $A_{\mu} -
\partial_{\mu} \phi$, where $\phi$ is the phase of the wave function
of Cooper pairs in the ground state of a superconductor).
To study the limit of large separations one needs the zero frequency
Matsubara term. Suppose $\Pi_{tr}(\omega=0, \vec{k}, T) = m_0^2(T) +
\gamma(T)\vec{k}^2 + O(\vec{k}^4)$. In this case we get
\begin{align}
r_{TM}(i\omega=0 , k) &= 1, \label{TM5}
\\
r_{TE}(i\omega=0, k) &= \frac{k_z^{(v2)} - k_z^{(m2)}}{k_z^{(v2)} +
k_z^{(m2)} } , \label{TE5}
\end{align}
where
\begin{align}
k_z^{(v2)} &= \sqrt{k_x^2 + k_y^2}, \\
k_z^{(m2)} &= \sqrt{m_0^2(T)/(1 + \gamma(T)) + k_x^2 + k_y^2} .
\end{align}
Note these expressions are valid for small $k$ only, nevertheless
for a nonzero $m_0(T)$ such behavior of $\Pi_{tr}(\omega=0, \vec{k},
T)$ for small $k$ leads to the ideal metal asymptotics of free
energy at separations $4\pi T a \gg 1$: $\mathcal{F} \sim -
T\zeta(3)/(8\pi a^2)$.


As follows from equation (\ref{e2}), results (\ref{TM5}) and
(\ref{TE5}) correspond to zero frequency reflection coefficients for
the electromagnetic wave reflecting from a {\it superconductor} (if
Drude model of the permittivity is valid for the normal component of
the two-fluid model of a superconductor, see also Refs.\cite{Bim1,
Bim2}). Indeed, the photon field becomes massive with a mass
$m_0(T)$ as follows from equation (\ref{e2}), the condition
$\Pi_{tr}(\omega=0, \vec{k})|_{|\vec{k}|\to 0} = m_0^2(T)$ is the
principal condition characterizing superconductivity. Equation
(\ref{e2}) and the property $i\omega \vec{B} = {\rm rot} \vec{E}$
yield the equation for a magnetic field with a mass $m_0(T)$,
Meissner effect immediately follows from the equation for a magnetic
field. Zero frequency reflection coefficients (\ref{TM5}),
(\ref{TE5}) coincide with zero frequency reflection coefficients for
the plasma model of permittivity if one puts formally $m_0^2(T)/(1 +
\gamma(T))=\omega_{pl}^2$. In fact, if one considers plasma model
description of metals at large separations (or high temperatures)
one should try to explain at the same time why superconductivity is
not present in the same system.



In my opinion, one of the principal questions in the Casimir effect
is to understand the conditions under which it is possible to use
the approximation (\ref{term2}) and the resulting equation
(\ref{main}) in the Casimir effect, i.e. when it is possible to
neglect  spatial dispersion. Spatial dispersion (i.e. dependence of
$\Pi^{mn}$ on the frequency $\omega$ and the wave vector $\vec{k}$)
should be taken into account in metals and it is important in metals
indeed, it yields Debye screening, which immediately follows from
$\varepsilon_l(\omega=0, \vec{k})$. Of course, spatial dispersion
makes transition from momentum space to coordinate space complicated
from the mathematical point of view. However, it is worth studying
the issue since the solution of the problem of high temperature
behavior of the Casimir force between  two metals can hardly be
obtained without a detailed study of spatial dispersion and
properties of the polarization operator at finite temperature.

\section{Fermions in a layer}

In this section we find reflection coefficients for transverse
electric (TE) and transverse magnetic (TM) modes reflecting from the
$2+1$ layer  with fermions at $z=0$. High temperature results for a
graphene - ideal metal system are discussed.

 Equations
\begin{equation}
\partial_\mu F^{\mu\nu} +\delta(z) \Pi^{\nu\rho}A_\rho =0 \quad
\end{equation}
lead to conditions
\begin{equation}
\partial_z A_m |_{z=+0} - \partial_z A_m |_{z=-0} = \Pi_{mn} A^n|_{z=0}
. \label{SQ}
\end{equation}

Let's consider the condition
\begin{equation}
\partial_0 A^0 + \partial_l A^l + \partial_p A^p = 0, \label{gauge}
\end{equation}
here a direction of the wave vector in the $xy$ plane is along the
coordinate $l$, $p$ is a transverse direction. In fact, the
condition (\ref{gauge}) is quite convenient for a description of
transverse electric and transverse magnetic modes of the propagating
electromagnetic wave.

For a nonzero $A_p$, the condition $\partial_p A_p=0$ and conditions
$A_l=A_z=A_0=0$ describe propagation of the TE electromagnetic wave
(the electric field is parallel to the surface $z=0$) since $E_p
\sim A_p$.

For the TE wave we have:
\begin{align}
A_p &= e^{i k l} e^{i k_z z} + r_{TE} e^{i k l}e^{-i k_z z} \quad
\text{for} \quad z<0  \label{TE11}
\\ A_p &= e^{i k_z z} e^{i k l} t_{TE} \quad \text{for} \quad z>0 \label{TE2}
\end{align}
and
\begin{equation}
\Pi_{p n} A^n = -\Pi_{pp}(\omega, k) A_p . \label{T1}
\end{equation}
Here $k_z^2=\omega^2-k^2$, $k^2=k_x^2+k_y^2$. From continuity of
potentials at $z=0$ one gets $1+r_{TE}=t_{TE}$. Now one substitutes
(\ref{TE11}) and (\ref{TE2}) into (\ref{SQ}) and uses (\ref{T1}) to
obtain:
\begin{equation}
r_{TE}(i\omega,k) = \frac{\Pi_{pp}(i\omega, k)}{2
\sqrt{\omega^2+k^2} - \Pi_{pp}(i\omega,k)}  = \frac{\Pi_{pp}(i
\omega,k)}{2\sqrt{\omega^2+ k^2}} \Biggl(1- \frac{\Pi_{pp}(i \omega,
k)}{2\sqrt{\omega^2+ k^2}}\Biggr)^{-1} . \label{rte2}
\end{equation}

Conditions $A_p=A_z=0$, $k_0 A_0=k A_l$ describe the transverse
magnetic (TM) wave. This choice of vector potentials describes TM
wave since $E_z \sim
\partial_z A_0 $ or $B_p \sim \partial_z A_l$.
For  $A_0$ we have:
\begin{align}
A_0 &= e^{i k l} e^{i k_z z} + r_{A_0} e^{i k l}e^{-i k_z z} \quad
\text{for} \quad z<0  \label{TM1}
\\ A_0 &= e^{i k_z z}e^{i k l} t_{A_0} \quad \text{for} \quad z>0 \label{TM2}.
\end{align}

Two conditions follow from gauge invariance:
\begin{align}
\omega \Pi_{00} (\omega,k) - k \Pi_{l0} (\omega,k) &= 0 ,\nonumber\\
\omega \Pi_{0l} (\omega,k) - k \Pi_{ll} (\omega,k) &= 0 , \nonumber
\end{align}
which yield
\begin{equation}
\omega^2 \Pi_{00}(\omega, k) = k^2 \Pi_{ll} (\omega, k). \label{ID1}
\end{equation}
Thus one gets
\begin{equation}
\Pi_{00}A^0 + \Pi_{0l} A^l = - \frac{\omega^2- k^2}{\omega^2}
\Pi_{ll} A_0 = - \frac{k_z^2}{\omega^2} \Pi_{ll} A_0 ,
\end{equation}
and from (\ref{SQ}) the condition
\begin{equation}
2i k_z (t_{A_0} + r_{A_0} -1) = - \frac{k_z^2}{\omega^2} \Pi_{ll}
A_0 (1+ r_{A_0})
\end{equation}
follows. After use of $r_{TM}(i\omega,k) = - r_{A_0}(i\omega,k)$ and
Wick rotation one obtains
\begin{equation}
r_{TM}(i\omega,k) =  - \frac{\sqrt{\omega^2 +k^2}}{2\omega^2}
\Pi_{ll}(i\omega, k) \Biggl(1 -
\frac{\sqrt{\omega^2+k^2}}{2\omega^2} \Pi_{ll}(i\omega,
k)\Biggr)^{-1} \label{rtm2}
\end{equation}

It is interesting to see how these results can be interpereted in
terms of diagrams. Consider first $2+1$ fermions interacting with an
atom.

\begin{figure}
\centering \includegraphics[width=7cm]{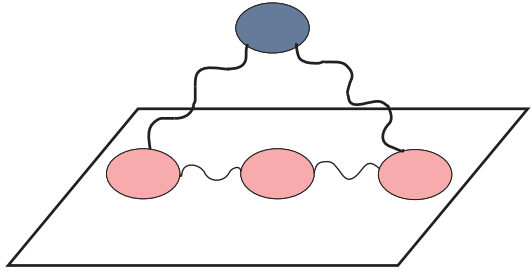}
\caption{Typical term in the Casimir-Polder energy between an atom
and fermions in a $2+1$ layer. Thick wavy line represents $D(a)$,
thin wavy line represents $D(0)$.} \label{Feynman1}
\end{figure}

In the gauge $A_0=0$ the longitudinal part of the free photon
propagator  has the form ($i=1,2$):
\begin{equation}
D^{L}_{ij} (i\omega, k, z) = \frac{k_i k_j}{k^2} \frac{
\sqrt{\omega^2 + k^2} e^{-|z|\sqrt{\omega^2 + k^2}} }{2 \omega^2} ,
\label{DL}
\end{equation}
the transverse part of the free photon propagator has the form:
\begin{equation}
D^{T}_{ij} (i\omega,k, z) = \Bigl(\delta_{ij} - \frac{k_i
k_j}{k^2}\Bigr) \frac{e^{-|z|\sqrt{\omega^2+ k^2}}}{2\sqrt{\omega^2+
k^2}} . \label{DT}
\end{equation}
Consider  TE part of the formula (\ref{an}). Substituting
(\ref{rte2}) into (\ref{an}) and using (\ref{DT}) we get:
\begin{align}
\Pi_{pp}^{at} D_{pp} &= \Pi_{pp}^{at} D^{T}_{pp}(2a)\cdot r_{TE}=
 \nonumber \\
&=\Pi_{pp}^{at} D^{T}_{pp}(a) \Bigl(\Pi_{pp} +
\Pi_{pp}D^{T}_{pp}(0)\Pi_{pp} + \ldots\Bigr) D^{T}_{pp}(a)
\label{expan}
\end{align}
The expression (\ref{expan}) has a clear diagrammatic representation
(one of the terms is shown on Fig.\ref{Feynman1}). Sum of the terms
in round parenthesis in (\ref{expan}) is the sum of RPA diagrams in
the two-dimensional layer (note that photon propagators do not
depend on $z$ there). Taking the trace of (\ref{expan}) one gets TE
contribution to the Casimir-Polder energy of interaction of the atom
and $2+1$ fermion layer.

\begin{figure} \centering \includegraphics[width=6cm]{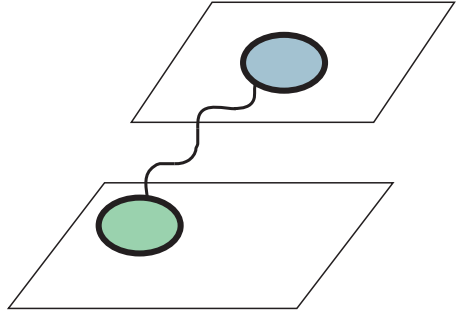}
\caption{Cluster in the Lifshitz formula. Thick ellipse represents
sum of RPA diagrams in a fermion layer. Thick wavy line represents
$D(a)$.} \label{Feynman2}
\end{figure}

\begin{figure}
\centering \includegraphics[width=10cm]{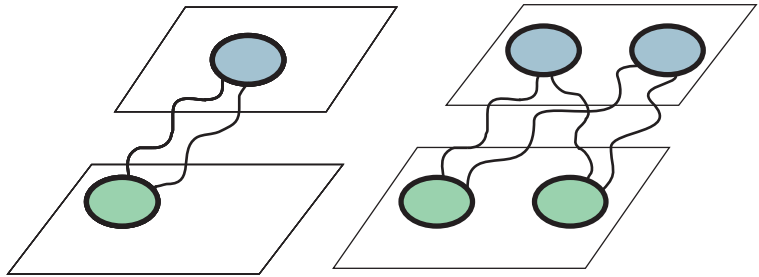}
\caption{Diagram representation of the Lifshitz free energy for two
parallel fermion layers} \label{Feynman3}
\end{figure}

Consider the formula (\ref{EL}) for two parallel layers with
fermions separated by a distance $a$. One gets diagrammatic
representation of free energy expanding the logarithm. The factor
$1/n$ is a standard factor arising in representation of
thermodynamic potential in terms of closed loop diagrams with $n$
equivalent clusters connected by $n$ photon lines. The cluster is
shown on Fig.\ref{Feynman2}, typical diagrams contributing to the
Lifshitz free energy are shown on Fig.\ref{Feynman3}. Recent
discussions of related multiple scattering techniques can be found
in Refs.\cite{EB}, \cite{Maghrebi}.

Graphene is a typical $2+1$ system that can be studied along the
lines described in this part.  Quasiparticles in graphene\cite{Geim}
obey a linear dispersion law $\omega= v_F k$ ($v_F\approx c/300$ is
a Fermi velocity, $c$ is a speed of light) at energies less than $2$
eV. There are $N=4$ species of fermions in graphene. The
polarization operator for $2+1$ fermions at finite temperature was
found in \cite{Mar1}.

Define ${\rm tr}\Pi\equiv\Pi_m^m$. Due to the property (\ref{ID1})
and
\begin{equation}
 {\rm tr}\Pi(i \omega, k) = \Pi_{00} (i \omega, k) \frac{\omega^2 +
k^2}{k^2} - \Pi_{pp} (i \omega, k)
\end{equation}
one can rewrite reflection coefficients (\ref{rte2}) and
(\ref{rtm2}) in the form given in \cite{Mar1}:
\begin{equation} r_{TM}(\omega, k) =\frac{k_z  \Pi_{00}}{k_z \Pi_{00}  + 2 i k^2},
\qquad r_{TE}(\omega, k)= - \frac{ k_z^2 \Pi_{00}+ k^2 {\rm tr}\Pi}
            {k_z^2 \Pi_{00} +  k^2 ({\rm tr}\Pi - 2 i k_z)}.
\label{rTETM-grPi}
\end{equation}

For $k \to 0$ one gets for the zero mass gap and zero chemical
potential ($\alpha$ is the coupling constant):
\begin{align}
&\Pi^{00}(i\omega=0, k) =  \frac{4 \alpha N T \ln2}{v_F^2}
+\frac{\alpha N k^2}{12 T} + \dots ,  \nonumber \\
 &{\rm tr}\Pi(i \omega=0, k) - \Pi^{00} (i \omega=0, k) =
  \frac{\alpha N v_F^2 k^2}{6T} + \dots . \nonumber
\end{align}
For the ideal metal $r_{TM}=1, r_{TE}=-1$ for all frequencies and
wave vectors. Zero Matsubara TM and TE terms yield following
high-temperature behavior of  free energy (\ref{EL}) in the graphene
-- ideal metal system:
\begin{align}
  {\mathcal{F}}_{0 \rm TM } &=
       -\frac{T\zeta(3)}{16 \pi a^2}  + \dots
               ,  \\
  {\mathcal F}_{0\rm TE}
         &=  -\frac{\a N v_F^2 }{192 \pi a^3} + \dots.
\end{align}
Here
\begin{equation}
-\frac{T\zeta(3)}{16 \pi a^2} \equiv {\mathcal F}_{\rm
Drude}\vert_{T\to\infty}=\frac12 {\mathcal F}_{\rm
id}\vert_{T\to\infty} .
\end{equation}
is the high-temperature  asymptotics of the metal -- metal system
with a Drude model of permittivity used \cite{Sernelius2} -
\cite{Drude1}, which is equal to one half of the high-temperature
asymptotics in the metal -- metal system with ideal boundary
conditions (\ref{ideal}) or the plasma model of permittivity used
\cite{Decca}. The zero frequency TE Matsubara term is suppressed by
a factor $\alpha N v_F^2$ and additional power of $1/(Ta)$.

The high-temperature asymptotics of free energy in a graphene-metal
system coincides with the Drude high-temperature asymptotics of the
metal-metal system. Detailed analysis \cite{Mar1} shows that the
high temperature behavior in the graphene-metal system takes place
already at separations of the order of $100$ nm at temperature
$T=300$K, which makes systems with graphene very promising for
experimental studies of the finite temperature Casimir effect.

The example of this section shows that nonlocality arising due to
interaction of photons via a polarization operator plays a crucial
role in the high temperature asymptotics of the system. Vacuum
effects arising due to fermions are essentially nonlocal and lead to
spatial dispersion. One of the most challenging problems in field
theory is to unify nonlocality in momentum space with boundary
problems in coordinate space. Casimir effect is a natural area for
further developments in this direction.

\section{Conclusions}
Casimir effect is an area of research where one applies and develops
methods of quantum field theory in the presence of a medium.
Throughout the paper I discuss several techniques which proved to be
efficient in the theory of the Casimir effect.


Casimir-Polder effect is studied in Feynman gauge of vector
potentials in Sec.$2$, the Casimir-Polder energy of an anisotropic
atom above a flat surface with boundary conditions of the ideal
conductor and above a dielectric surface is found. Scattering
approach is an efficient method in the theory of the Casimir effect,
it is introduced in Sec.$3$ and applied to flat and periodic
geometries separated by a vacuum slit.

Properties of the medium determine important characteristics of
Casimir systems such as the high temperature asymptotics of free
energy once one knows dielectric permittivity of the medium. An
alternative approach is to consider properties of the medium through
evaluation of components of the polarization operator, in the
current paper this idea is being developed in Sections $4$ and $5$.
In Sec.$5$ exceptional finite temperature properties of systems with
graphene are discussed.

 \vspace{0.5cm} {\noindent \it Further reading}

A discussion of interaction in the Casimir effect is given in the
paper \cite{Jaffe}. In the review \cite{Barash} applications of the
argument principle to flat geometries are considered, dissipation
and dispersion are discussed. Papers \cite{Nesterenko, Bordag5}
yield detailed derivation of the Lifshitz result for the plasma
model of permittivity. Finite temperature Casimir effect is reviewed
in \cite{Brevik7}, see also a review \cite{rev15}. Examples of
scattering theory applied to spherical and cylindrical geometries
can be found  in papers \cite{J1} - \cite{Pirozhenko} and in Chapter
$10$ of the book \cite{rev4}. Casimir repulsion due to Chern-Simons
term was studied in Refs.\cite{Vas5, Pismak1}, scattering of
electromagnetic waves on a plane with Chern-Simons term was
considered in Ref.\cite{Pismak2}. Casimir-Polder interaction and
macroscopic QED are reviewed in \cite{rev16, rev17, Parsegian},
scattering approach to Casimir-Polder effect is discussed in
Ref.\cite{French5}. Recent advances in microstructured geometries
and Casimir effect are summarized in a review \cite{rev19}.
Techniques for evaluation of functional determinants are discussed
in \cite{Dowker}, methods of heat kernel and zeta function are
presented in reviews \cite{Santangelo, rev20} and books
 \cite{rev10}, \cite{rev11}, \cite{FV}, \cite{rev12} in detail.
One can find examples of zeta function technique in the Casimir
effect problems in papers \cite{pap1} - \cite{pap6}.



\end{document}